\documentclass{PoS}

\usepackage{amsmath,amsfonts,amssymb}

\usepackage{graphicx}
\usepackage{subcaption}

\newcommand\be{\begin{equation}}
\newcommand\ee{\end{equation}}
\newcommand\bea{\begin{eqnarray}}
\newcommand\eea{\end{eqnarray}}

\newcommand\s{s}
\newcommand\kk{\omega}
  
\newcommand\ka{\Lambda}

   \def\>#1{{\mathbf#1}}

\title{Generalized noncommutative Snyder spaces and projective geometry}

\ShortTitle{Generalized noncommutative Snyder spaces}

\author{\speaker{Giulia Gubitosi}\\
        Departamento de F\'isica, Universidad de Burgos, 09001 Burgos, Spain\\
        E-mail: \email{giulia.gubitosi@gmail.com}}

\author{Angel Ballesteros\\
       Departamento de F\'isica, Universidad de Burgos, 09001 Burgos, Spain\\
        E-mail: \email{angelb@ubu.es}}
        
 \author{Francisco J. Herranz\\
        Departamento de F\'isica, Universidad de Burgos, 09001 Burgos, Spain\\
        E-mail: \email{fjherranz@ubu.es}}

\abstract{Given a group of kinematical symmetry generators, one can construct a compatible noncommutative spacetime and deformed phase space  by means of projective geometry. This was the main idea behind the very first model of noncommutative spacetime, proposed by H.S.~Snyder in 1947. In this framework, spacetime coordinates are the translation generators over a manifold that is symmetric under the required generators, while momenta are  projective coordinates on such a manifold. In these proceedings we review the construction of Euclidean   and  Lorentzian noncommutative Snyder spaces and investigate the freedom left by this construction in the choice of the physical momenta, because of different available choices of projective coordinates. In particular, we derive a quasi-canonical structure for both the Euclidean and Lorentzian Snyder noncommutative models such that  their phase space algebra is diagonal  although no longer quadratic.}

\FullConference{Corfu Summer Institute 2019 "School and Workshops on Elementary Particle Physics and Gravity" (CORFU2019)\\
		31 August - 25 September 2019\\
		Corfu, Greece}

\begin{document}

\section{Introduction}

Noncommutative spacetimes play an important role in quantum gravity research, since they provide a way to formalize the ``foamy'' or fuzzy features that spacetime is thought to acquire at the Planck scale  \cite{AmelinoCamelia:2008qg,Garay:1994en, Hossenfelder:2012jw}. Somewhat related to this, 
another possibility that has been long entertained in quantum gravity research is that of the emergence of a minimal length uncertainty and the associated generalised uncertainty principle (GUP)  \cite{Garay:1994en, Amati:1988tn, Konishi:1989wk,  Kempf:1994su,Quesne:2006is}.

It is interesting to investigate whether such quantum features can be introduced without spoiling invariance under relativistic symmetries and how they are related to each other. This was the driving idea of H.S.~Snyder's work \cite{Snyder1947} in 1947: he was able to show that one can construct a noncommutative spacetime (and correspondingly deformed phase space leading to a GUP \cite{Quesne:2006is, Mignemi:2009ji, Amelino-Camelia:2018wzu}) that is invariant under the usual special-relativistic Lorentz  symmetries. 
The construction assumes that momentum space is a de Sitter manifold, spacetime coordinates are identified with the translation generators on this manifold (so that their noncommutativity is a consequence of the manifold's curvature), and physical momenta are taken to be the Beltrami projective coordinates on the manifold. Physical momenta thus constructed satisfy two requirements: they are commutative and transform classically under Lorentz symmetries.

In recent work \cite{BGH2019} we extended the construction to  noncommutative spacetimes that are compatible with Galilean and  Carrollian relativistic symmetries. 
There, we noticed that there is some freedom in the choice of the physical momenta, since different projective coordinates on the momentum manifold satisfy the two requirements mentioned above. This freedom is  very relevant for the study of the phenomenological consequences of the Snyder model \cite{Amelino-Camelia:2018wzu, Nozari:2015iba,  Mignemi:2011gr, Lu:2011it,Ivetic:2015cwa,Mignemi:2011wh,Mignemi:2013aua,Banerjee:2006wf,Ivetic:2018}, since different choices of momenta and resulting different phase spaces would lead to different physical predictions.

In these proceedings we investigate in depth the issue of the freedom in the choice of momenta, focussing on Snyder models with Euclidean and Lorentzian symmetries. We expose  different deformed phase spaces that arise from different choices of physical momenta.  We find that any choice of projective coordinates on the curved manifold gives a viable choice of momenta. Moreover, we derive
a quasi-canonical structure for both the Euclidean and Lorentzian Snyder noncommutative models such that  their phase space algebra is diagonal  although no longer quadratic. We show that the momenta associated to this choice can be identified with the ambient coordinates of the curved manifold.


\section{Projective geometry construction of the Snyder--Euclidean model}

We start by reviewing  the construction of the two-dimensional (2D) Snyder--Euclidean noncommutative space, emphasizing the different possible choices of spatial momenta and their implications.

\begin{figure}[htbp]
\begin{center}
\begin{subfigure}[b]{0.42\linewidth}
    \includegraphics[width=\linewidth]{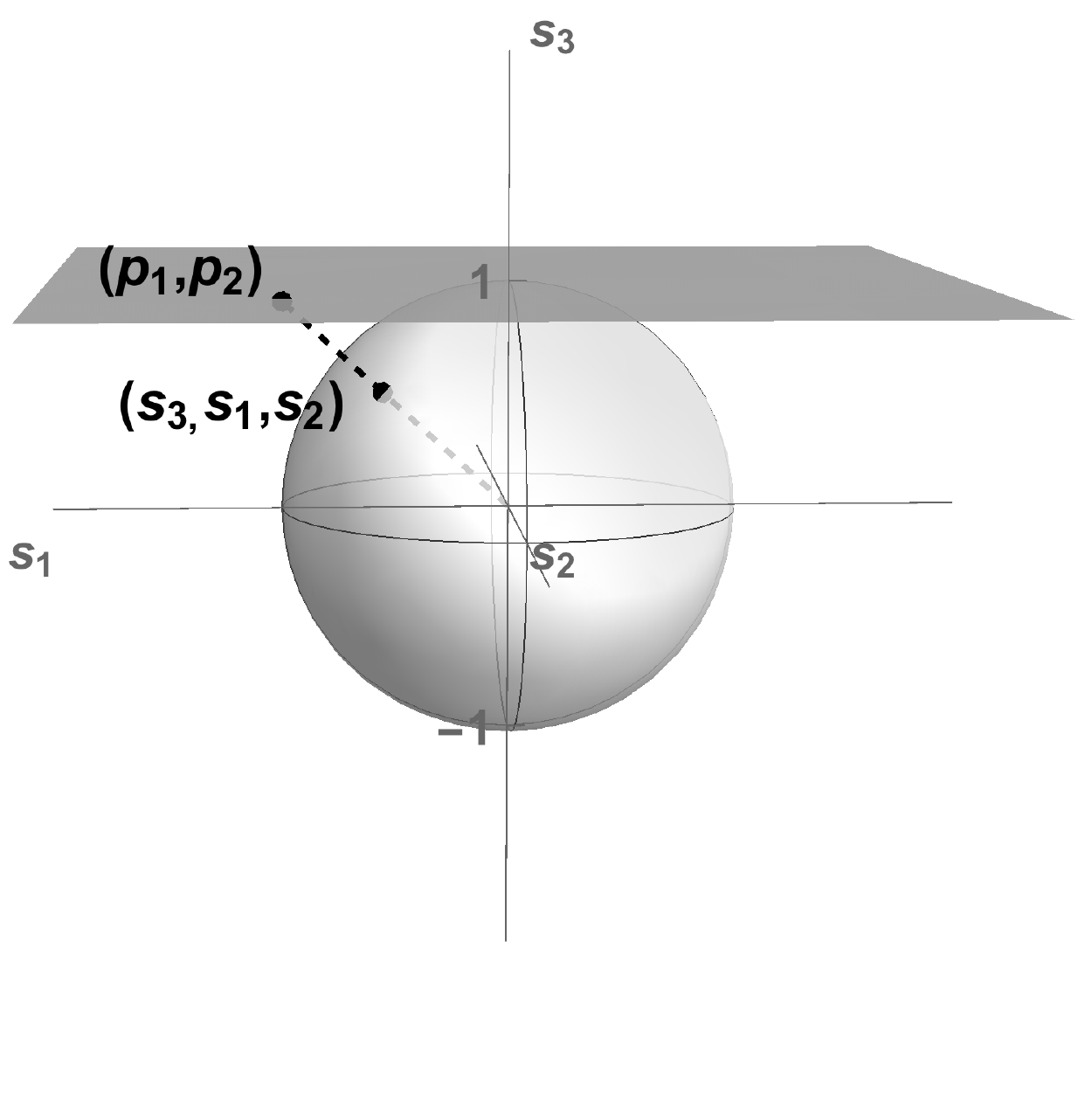}
  \end{subfigure}
  \quad
  \begin{subfigure}[b]{0.40\linewidth}
    \includegraphics[width=\linewidth]{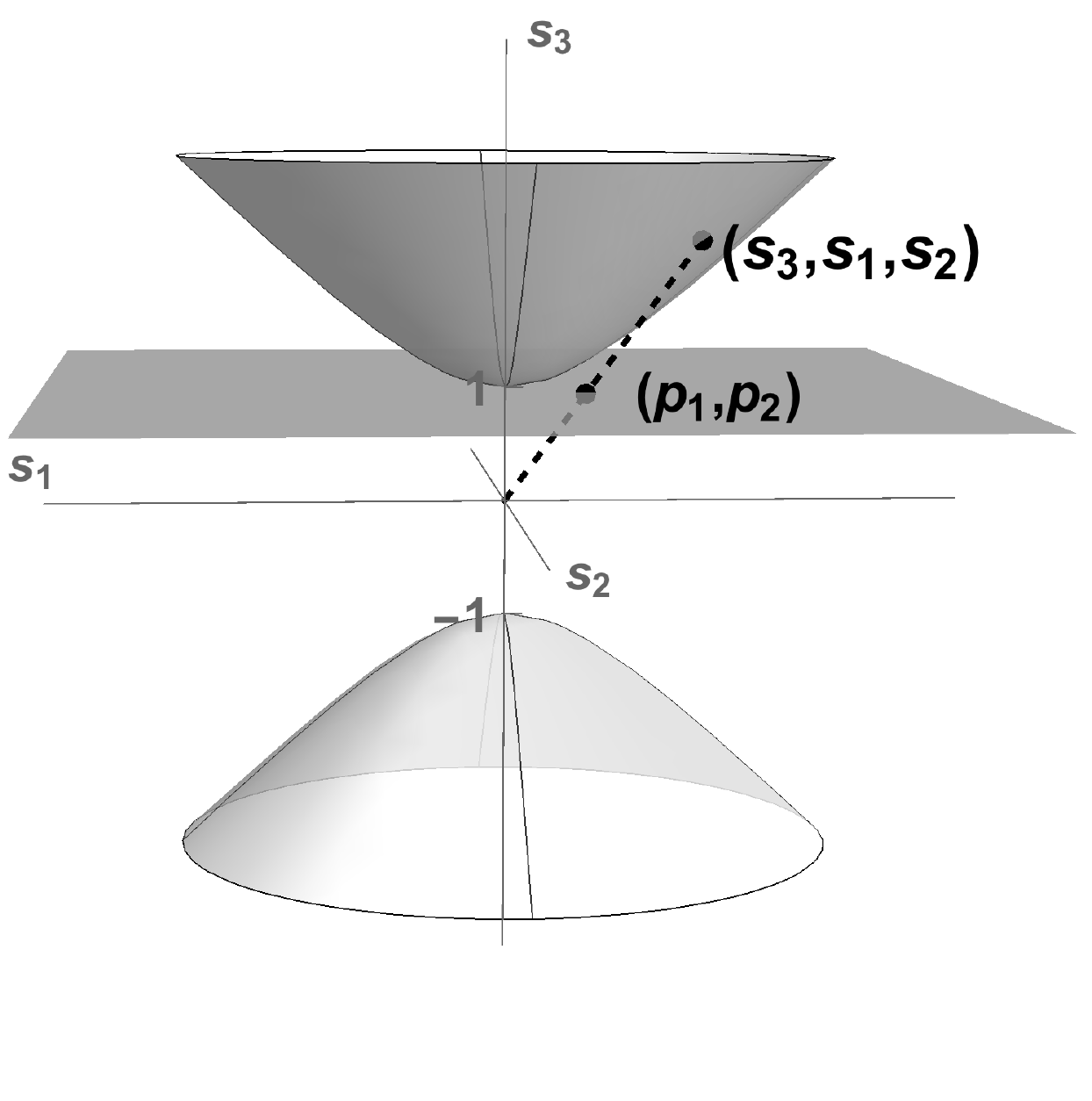}
  \end{subfigure}
\caption{Beltrami projective coordinates for  the sphere on the left and the hyperboloid on the right.}
\label{fig:euclidean}
\end{center}
\end{figure}

Since we require that the Snyder--Euclidean noncommutative space is invariant under rotations, we take spatial coordinates to be the translation generators over a 2D  manifold with constant Gaussian curvature $\omega$ that is either a sphere ($\omega>0$) or a two-sheeted hyperboloid ($\omega<0$), and enjoys ${\rm SO}(3)$ or ${\rm SO}(2,1)$ symmetry, respectively. Specifically, the Lie algebra of isometries of the manifold is generated by the   rotation $J$ and translations $P_{1},P_{2}$,  with commutation relations 
\begin{equation}
 [J ,P_{1}]=   P_2, \qquad  [J ,P_{2}]= -P_1, \qquad [P_{1},P_{2}]=\omega\, J. \label{ca}
 \end{equation}
As we showed in \cite{BGH2019}, the  homogeneous spaces of the Lie groups with Lie algebras (\ref{ca}) (i.e.~the curved manifolds) can be described in terms of embedding coordinates $(s_{3},s_{1},s_{2})$ that satisfy the following constraint
\begin{equation}
\Sigma_{\omega}\ :\ s_{3}^{2}+\omega \left(s_{1}^{2}+s_{2}^{2}\right)=1.
\label{sigma}
\end{equation}
In terms of these ambient coordinates, the vector fields corresponding to the symmetry transformations read
\be
P_{1}=\s_3 \,\frac{\partial}{\partial \s_1}   - \kk \, \s_1\, \frac{\partial}{\partial \s_3}   ,\qquad
P_{2}=  \s_3 \, \frac{\partial}{\partial \s_2}- \kk  \,\s_2 \, \frac{\partial}{\partial \s_3}    ,\qquad 
J =  \s_2 \, \frac{\partial}{\partial \s_1}    -\s_1 \,\frac{\partial}{\partial \s_2}  .
\label{ck}
\ee
 The Snyder--Euclidean noncommutative space is then obtained upon the identification
\be
x_{1}:=P_{1},\qquad x_{2}:=P_{2},
\ee
so that   spatial coordinates inherit the noncommutativity from the curvature of the underlying manifold:
\be
\left[x_{1},x_{2}\right]=\omega\, J.
\ee
In order to construct a full phase space we define momenta as objects that are commutative and transform as vector under rotations. It turns out that several kinds of projective coordinates of the curved manifold satisfy these requirements \cite{BGH2019}. The ones corresponding to the original choice by Snyder \cite{Snyder1947} are  the Beltrami projective coordinates:
\be
p_{1}:=\frac{s_{1}}{s_{3}}, \qquad p_{2}:=\frac{s_{2}}{s_{3}}.
\label{momenta1}
\ee
The coordinates $(p_1,p_2)$ are obtained as   
 the central stereographic projection with pole  
$(0,0,0)\in \mathbb R^{3}$ of a point $ Q=(s_3, s_1,s_2)\in \Sigma_{\omega}$  (\ref{sigma})  onto   the projective plane with $s_3=1$ as depicted in figure~\ref{fig:euclidean}. This establishes the relations $(i=1,2)$
\begin{equation}
s_3=\frac{1}{\sqrt{1+ \kk (p_1^2+p_2^2)}},\qquad
s_i=\frac{p_i}{\sqrt{1+ \omega (p_1^2+p_2^2)}},  \qquad  p_i=\frac{s_i}{s_3}  .
\label{da}
\end{equation}

 Thus the   origin  $O=(1,0,0)\in \Sigma_\omega$ projects to the origin $(p_1,p_2)=(0,0)$ in the projective space.  The  domain of $(p_1,p_2)$   depends on the   sign of the curvature $\omega$ since the equations \eqref{da} require that
\be
1+ \omega (p_1^2+p_2^2)>0 .
\label{condition}
\ee
In particular we find that~\cite{BaBlHeMu14}:
\begin{itemize}
\item  When the underlying curved manifold is a  sphere, corresponding to the Snyder--Euclidean model with  $\kk>0$,    the condition (\ref{condition}) is always satisfied, so that there is no restriction on the momentum domain
\be
p_i\in(-\infty,+\infty)  .
\label{domainS}
\ee
This corresponds to the fact that the projection sends the points in the equator in $ \Sigma_\kk$  (\ref{sigma}), characterized by   $\s_3=0$   and $\s_1^2+\s_2^2=1/\kk$,   to infinity in the projective plane,  so that the projection (\ref{da}) is   well-defined for the whole  hemisphere  with $\s_3>0$.

\item
  When the underlying curved manifold is a hyperbolic space (the   sheet of the hyperboloid with $\s_3\ge 1$), corresponding to the Snyder--Euclidean model with $\kk<0$,  the momentum space domain is restricted, since the condition (\ref{condition}) is only satisfied for 
 \be
p_1^2+p_2^2< \frac1{|\kk|} .\label{domainH}
 \ee
 Hence the momentum space can be identified with   the interior of a Klein disk  with   radius $1/\sqrt{|\kk|}$.
In fact,  the projection (\ref{da}) sends the points at the infinity in the upper sheet of the hyperboloid  with  $\s_3\to\infty$    to the circle
$ p_1^2+p_2^2 =  \frac1{|\kk|} $ (that is, the boundary of the Klein disk).     
 \end{itemize}

With the choice of  the physical momenta (\ref{momenta1}) one obtains the following phase space~\cite{BGH2019}\footnote{As we discussed in \cite{BGH2019}, starting from this phase space algebra one can easily define coordinates and momenta as quantum Hermitian operators. We refer the reader to \cite{BGH2019} for further details, since this issue is not relevant for the scopes of these notes.}:
\be
\begin{array}{ll}
\left[x_1,x_2\right]=  \kk\, J ,& \qquad \left[p_1,p_2\right]= 0,  \\[4pt]
\left[x_1,p_1\right]=1+\kk  \,p_1^2,& \qquad  \left[x_1,p_2\right]= \kk  \, p_1 p_2, \\[4pt]
 \left[x_2,p_2\right]= 1+\kk   \,p_2^2, &\qquad \left[x_2,p_1\right]= \kk  \, p_1 p_2 ,
\end{array}
\label{Snyder2Db}
\ee
where the rotation generator $J$  becomes the angular momentum in \eqref{Snyder2Db} given by
\be
 J= x_1\, p_2-x_2\, p_1 .
\ee
One can verify by direct computation that these phase space variables  satisfy the Jacobi identities. 
 Moreover,  both the space coordinates $x_{i}$ and momenta $p_{i}$ are transformed as vectors under  $J$:
 \be
 [ J, x_{1}]=x_{2},\qquad  [ J, x_{2}]=-  x_{1},\qquad  [ J, p_1]= p_2,\qquad  [ J, p_2]=-  p_1 ,
 \label{vector}
 \ee
 so that    the phase space (\ref{Snyder2Db}) is invariant under rotations, as required. 
 

 \subsection{Different choices of  momenta in the Snyder--Euclidean model}
 \label{21}

\begin{figure}[htbp]
\begin{center}
\begin{subfigure}[b]{0.41\linewidth}
    \includegraphics[width=\linewidth]{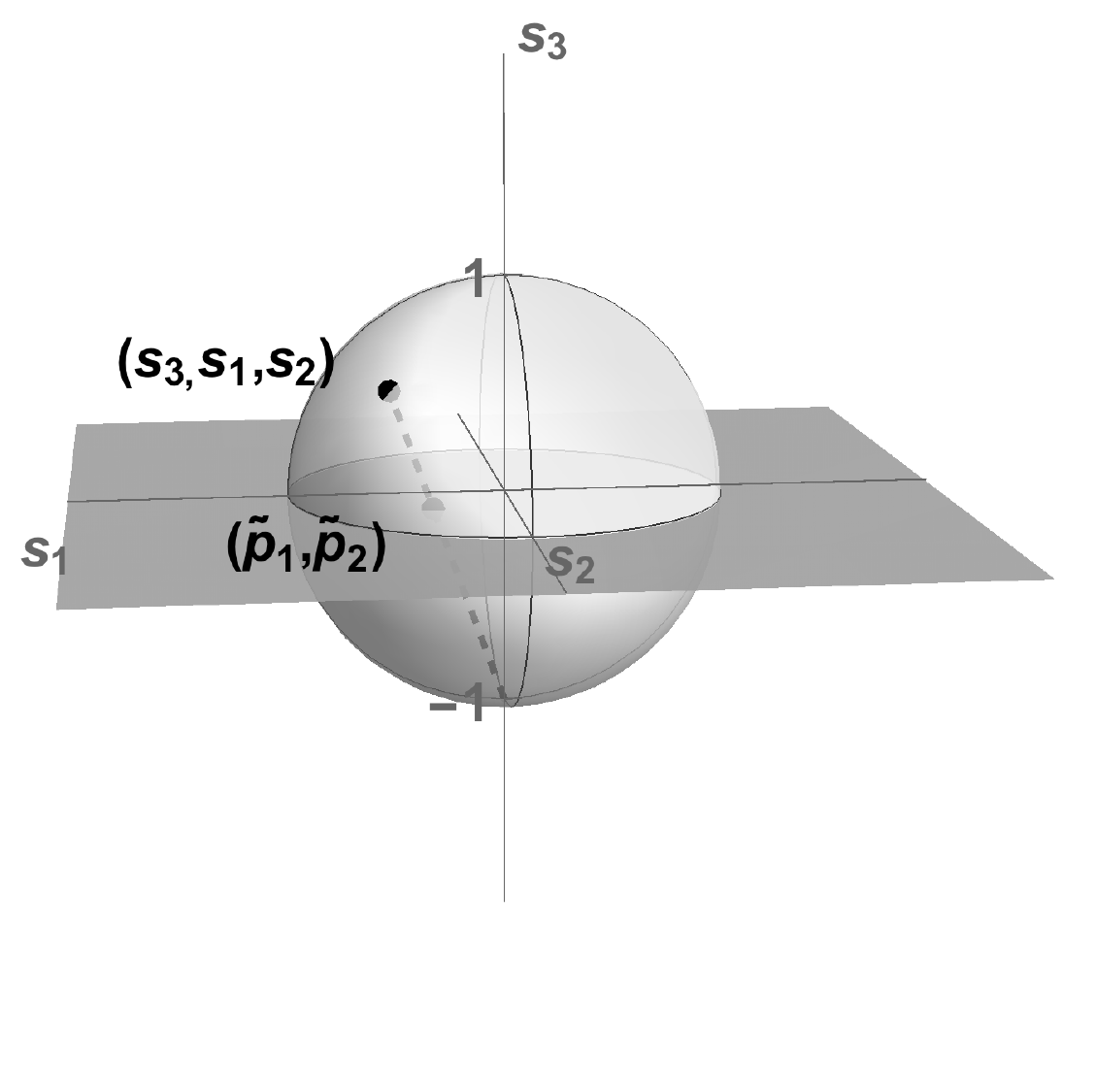}
  \end{subfigure}
  \quad
  \begin{subfigure}[b]{0.44\linewidth}
    \includegraphics[width=\linewidth]{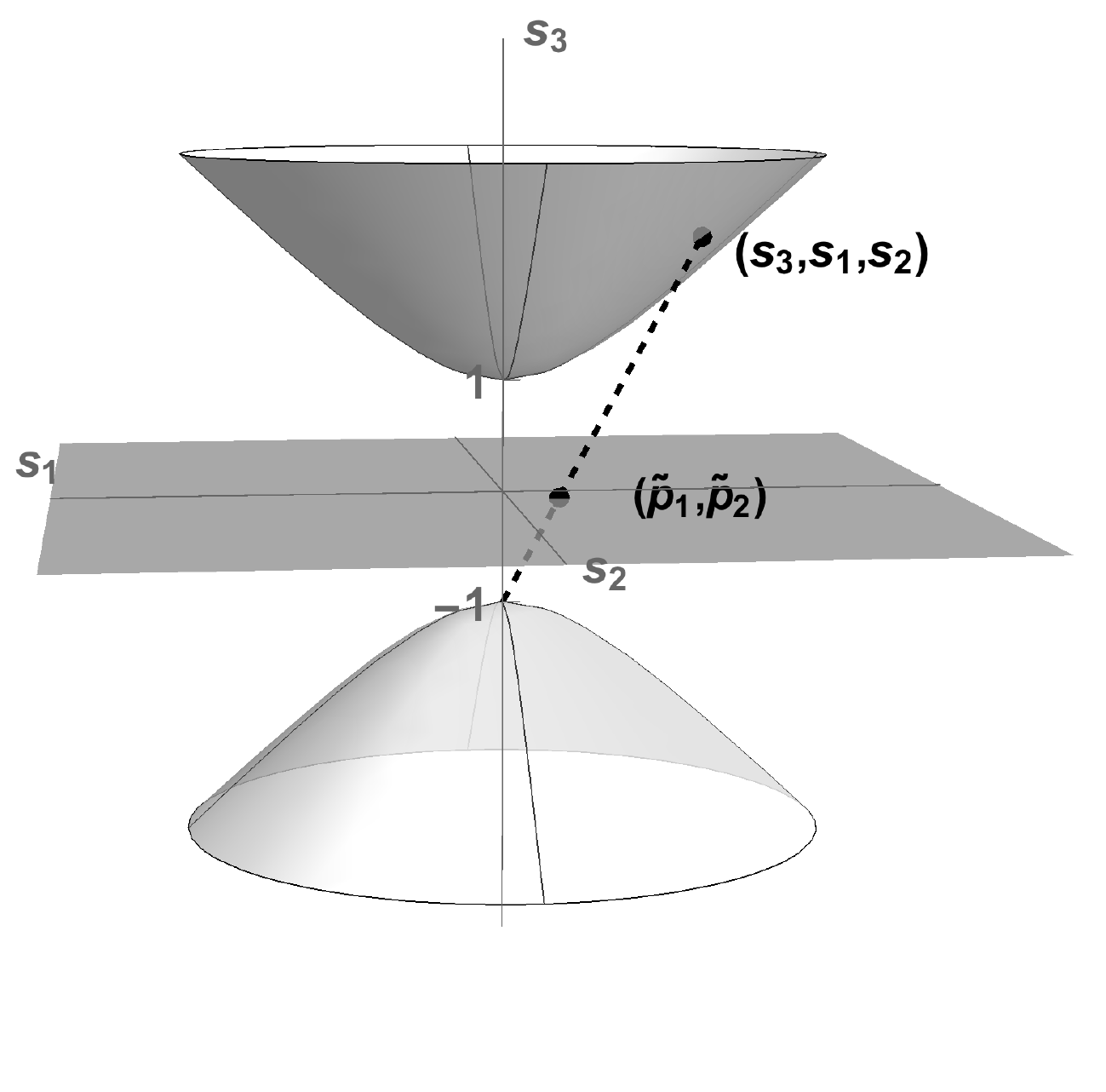}
  \end{subfigure}
\caption{Poincar\'e projective coordinates for  the sphere on the left and the hyperboloid on the right.}
\label{fig:euclideanP}
\end{center}
\end{figure}

As already noticed in \cite{Snyder1947}, the choice of physical momenta is not univocally determined by the requirements of being commutative and transforming as vectors under rotations. In fact, these requirements by themselves  do not completely fix the functional dependence of  momenta on the  embedding coordinates $(s_{3},s_{1},s_{2})$. In particular, the Beltrami projective coordinates are not the only available option. For example, one could identify momenta with the Poincar\'e projective coordinates $(\tilde p_1,\tilde p_2)$,  which come from 
 the stereographic projection with   pole  $(-1,0,0)\in \mathbb R^{3}$ onto   the projective plane with $s_3=0$ as shown in figure \ref{fig:euclideanP}. In this case the relation between the ambient and the projective coordinates reads $(i=1,2)$

\be
\s_3= \frac{1- \kk  (\tilde p_1^2+\tilde p_2^2)}{ {1+ \kk (\tilde p_1^2+\tilde p_2^2)}}   ,\qquad 
\s_i=\frac{2{\tilde p_i}}{ {1+ \kk (\tilde p_1^2+\tilde p_2^2) }} ,\qquad    {\tilde p_i}=\frac{\s_i}{1+\s_3} .
 \label{Poinc}
 \ee

This projection is well-defined for any point $Q=(s_3, s_1,s_2)\in \Sigma_\kk$  (\ref{sigma})  except for the pole $(-1,0,0)$ which projects to the infinity in both the sphere and in the hyperbolic space. The   origin  $O=(1,0,0)\in \Sigma_\omega$ again projects to the origin $(\tilde p_1,\tilde p_2)=(0,0)$ in the projective space. In order to deduce  the  domain of the Poincar\'e coordinates $(\tilde p_1,\tilde p_2)$ we shall make use of the following expression coming from (\ref{Poinc}):
\be
\tilde p_1^2+\tilde p_2^2=\frac 1\kk\, \frac{1-\s_3}{1+\s_3} .
\label{domP}
\ee
Thus we find that~\cite{BaBlHeMu14}: 
\begin{itemize}

\item  On the  sphere   with  $\kk>0$, it is verified that  $-1<\s_3\le 1$  so that  the relation (\ref{domP})  gives 
\be
\tilde p_i\in(-\infty,+\infty)  .
\label{domainS2}
\ee
  
 \item
 On the hyperbolic   case  with $\kk<0$ and $\s_3\ge 1$, the condition (\ref{domP}) leads to a restriction of the domain
   \be
\tilde p_1^2+\tilde p_2^2=\frac 1{|\kk|}\, \frac{ \s_3-1}{\s_3+1} < \frac1{|\kk|} ,
 \ee
 which can   be identified with the   interior of a  Poincar\'e disk (or conformal disk model)   with   radius $1/\sqrt{|\kk|}$, whose boundary corresponds to the points with $\s_3\to\infty$.
  \end{itemize}

Taking   the Poincar\'e coordinates as physical momenta and identifying space coordinates  with the translations over the curved manifold as done before, one obtains a new Snyder--Euclidean phase space  given by
\be
\begin{array}{ll}
\left[  x_1, x_2\right]= \kk \,  J ,& \qquad \left[\tilde p_1,\tilde p_2\right]= 0,  \\[4pt]
\left[ x_1,\tilde p_1\right]=\frac 12 \bigl( 1+\kk (\tilde p_1^2-\tilde p_2^2) \bigr),& \qquad  \left[ x_1,\tilde p_2\right]= \kk\, \tilde p_1 \tilde p_2, \\[4pt]
 \left[ x_2,\tilde p_2\right]=\frac 12 \bigl( 1+\kk (\tilde p_2^2-\tilde p_1^2) \bigr), &\qquad \left[ x_2,\tilde p_1\right]= \kk\, \tilde p_1 \tilde p_2 ,
\end{array}
\label{Snyder2Dp}
\ee
where the angular momentum  $  J$ is now represented by
\be
  J=  2 \, \frac{ x_1  \tilde p_2- x_2  \tilde p_1}{1-\kk (\tilde p_1^2+\tilde p_2^2)}   .
  \label{Jpoinc}
\ee
Notice that the above expression is consistently written since
\be
\bigl[ x_1  \tilde p_2- x_2 \tilde p_1, \tilde p_1^2+\tilde p_2^2 \bigr] =0.
\ee

The phase space (\ref{Snyder2Dp})  is viable in the sense that its variables  satisfy the Jacobi identities and the invariance under rotations is retained, that is,   momenta $\tilde p_{i}$ transform as vectors under   $  J$ in the same form given in (\ref{vector}).

Thus the expressions (\ref{Snyder2Dp})   provide another way to define  2D  Snyder--Euclidean models. Moreover, the two models    (\ref{Snyder2Db}) and (\ref{Snyder2Dp})   are related by the  following non-linear  change of momenta $(i=1,2)$:
\be
 \tilde p_i=\frac{p_i}{1+\sqrt{1+\kk (p_1^2+p_2^2)}},\qquad   p_i=\frac{2\tilde p_i}{1- {\kk (\tilde p_1^2+\tilde p_2^2)}} .
 \label{change}
\ee
Recall that    momenta  commute and notice also that the above map does not involve the  space coordinates, so preserving the 
noncommutative Snyder space $[x_{1},x_{2}]= \kk\, J $, as it should be, although the explicit representation for the angular momentum $J$ changes.

We remark  that other types of projections (choosing another pole and  plane $\s_3=$ constant) would lead to different explicit forms for the 2D Snyder--Euclidean models, keeping the  space coordinates but with different momenta. These possibilities would be related among themselves through transformations of the momenta similarly to (\ref{change}).


\subsection{Quasi-canonical structure of the Snyder--Euclidean model}
  \label{22}

 By taking into account the above results,   it is rather natural to analyse whether the  structure of the quadratic algebra (\ref{Snyder2Db})   can  further  be simplified.  And actually it is possible to choose momenta in such a manner that the phase space algebra becomes diagonal. Indeed, if we     apply the change of momenta given by 
\be
 \pi_i=\frac{p_i}{ \sqrt{ 1+\kk (p_1^2+p_2^2)} },\qquad p_i=\frac{\pi_i}{ \sqrt{ 1-\kk (\pi_1^2+\pi_2^2)} } ,
\label{map}
\ee
     the commutators  (\ref{Snyder2Db}) reduce to
  \be
\begin{array}{l}
[x_{1},x_{2}]= \kk\, J , \qquad
 [\pi_1,\pi_2]= 0,  \\[4pt]
[x_{i},\pi_j]= \delta_{ij} \sqrt{ 1-\kk (\pi_1^2+\pi_2^2)} \, .
\end{array}
\label{Snyder2Dqb}
\ee
As a consequence, the algebra (\ref{Snyder2Dqb}) is no longer quadratic but conveys commutativity of the non-diagonal terms, 
$ [x_{1},\pi_2]= [x_{2},\pi_1]=0$. This diagonal algebra was originally considered in \cite{Maggiore:1993kv} and then linked to the Snyder model in \cite{Battisti:2010sr, Ivetic:2018}.

Furthermore, the diagonal brackets are deformed in terms of the curvature parameter $\kk$ in a symmetrical manner governed by $\pi_1^2+\pi_2^2$. The  power series expansion in terms of $\kk$ turns out to be
\be
[x_{1},\pi_1]= [x_{2},\pi_2]=  \sqrt{ 1-\kk (\pi_1^2+\pi_2^2)} =  \left(1-  \frac 12 \kk  (\pi_1^2+\pi_2^2) \right)+ o[\kk^2] ,
\ee
hence for a small value of $\kk$, the Snyder--Euclidean phase space (\ref{Snyder2Dqb}) can be taken as a quadratic algebra.

The angular momentum $J$ adopts the    non-standard representation 
 \be
J= \frac  {x_{1}\, \pi_2-x_{2}\, \pi_1}{ \sqrt{ 1-\kk (\pi_1^2+\pi_2^2)} } , 
\label{Snyder2DJb}
\ee
 as in  Poincar\'e variables (\ref{Jpoinc}), where we note that
\be
\bigl[ x_{1}\, \pi_2-x_{2}\, \pi_1, \pi_1^2+\pi_2^2 \bigr] =0.
\ee
Still, the new momenta defined in  \eqref{map} satisfy the required properties: they are commutative and  transform as vectors under rotations. This can easily be  checked by explicitly verifying that the same commutators as in (\ref{vector}) hold for the new momenta $(\pi_1,\pi_2)$  and $J$.

The domain for the new momenta $(\pi_1,\pi_2)$  is of course different from that of $(p_{1},p_{2})$, but it is so in a consistent way, since the map (\ref{map}) implies the identity
\be
1+\kk (p_{1}^2+p_{2}^2) = \frac{1}{ 1-\kk (\pi_1^2+\pi_2^2)},
\ee
 thus leading to the   constraint (see (\ref{condition}))
\be
1+ \kk (p_{1}^2+p_{2}^2)>0 \quad \Longleftrightarrow \quad 1- \kk (\pi_1^2+\pi_2^2)>0 .
\ee
Therefore the domain of the new momentum space is interchanged between the spherical and   hyperbolic cases, that is
\be
\begin{array}{ll}
\mbox{Snyder--Euclidean space with $\kk>0$:}\quad
&\pi_1,\pi_2    \in  \bigl(-1/\sqrt{|\kk|}, + 1/\sqrt{|\kk|} \,\bigr)  . \\[2pt]
\mbox{Snyder--Euclidean space with $\kk<0$:}\quad  & \pi_1,\pi_2\in (-\infty,+\infty).  
 \end{array}
\label{domain2Db}
\ee

Since we have previously seen that the momenta $p_{i}$ and  $\tilde p_{i}$ have a geometrical interpretation as projective coordinates on the curved momentum manifold, it is natural to wonder about the geometrical role of this alternative choice $\pi_{i}$. By comparison between the maps \eqref{da}
 and \eqref{map} it is immediate to see that the momenta $\pi_{i}$ have a direct relation with the momentum space ambient coordinates $s_{i}$:
\be
 \s_1 \equiv \pi_1 , \qquad \s_2 \equiv  \pi_2,\qquad \s_3 \equiv \sqrt{ 1-\kk (\pi_1^2+\pi_2^2)} .\label{EuclideanAmbientMomenta}
 \ee
From this viewpoint the ambient coordinates $(\s_1,\s_2)$ are just the momenta 
$(\pi_1,\pi_2)$, while  the third  coordinate $\s_3$ determines the geodesic distance, say $p_r$, between the origin $O=(1,0,0)$ and a    point  $Q=(\s_3,\s_1,\s_2)\in \Sigma_\kk$  (\ref{sigma})   in the 3D ambient space (see~\cite{BaBlHeMu14,conf}), namely 
 \be
 \s_3=\cos(\sqrt{\kk}\, p_r) .
  \ee
This is just the term deforming the canonical commutators $[x_{i},\pi_i]$.
 
It is worth remarking that the generalization of   all the results  presented in this section   to arbitrary dimension $N$, so with ${\rm SO}(N+1)$ and ${\rm SO}(N,1)$ symmetry, is straightforward. In fact, the necessary tools, such as the $(N+1)$ ambient coordinates along with  the $N$ Beltrami and Poincar\'e coordinates on the $N$D spherical and hyperbolic spaces in terms of their constant (sectional) curvature, can be found in~\cite{Kepler}.


\section{Projective geometry construction of the  Snyder--Lorentzian model}

Similar issues as the ones exposed for the Snyder--Euclidean model also emerge  in the kinematical cases covering Snyder--Lorentzian, Snyder--Galilean and Snyder--Carrollian models which   were developed in \cite{BGH2019}. In these notes we focus on the Snyder--Lorentzian model, since the corresponding momentum choices for the Galilean and Carrollian cases can be deduced  by following the limiting procedures described in~\cite{BGH2019}.

Invariance under Lorentzian symmetries is achieved by taking the (3+1)D spacetime coordinates to be the translations over a (3+1)D manifold with constant sectional curvature equal to $-\Lambda$, where $\Lambda$ is the cosmological constant. This is either de Sitter  ($\Lambda>0$) or Anti-de Sitter ($\Lambda<0$) manifold, enjoying $ {\rm SO}(4,1)$ and ${\rm SO}(3,2)$ symmetry, respectively. The algebra of isometries is generated by Lorentzian boosts $K_{i}$, rotations $J_{i}$ and spacetime translations $P_{\alpha}$ which fulfil the following commutation rules
\be
\begin{array}{lll}
[J_i,J_j]=\epsilon_{ijk}J_k ,& \qquad [J_i,P_j]=\epsilon_{ijk}P_k , &\qquad
[J_i,K_j]=\epsilon_{ijk}K_k , \\[4pt]
\displaystyle{
  [K_i,P_0]=P_i  } , &\qquad {[K_i,P_j]=\frac 1{c^2}\, \delta_{ij}  P_0} ,    &\qquad {[K_i,K_j]=-\frac 1{c^2}\,\epsilon_{ijk} J_k} , 
\\[4pt][P_0,P_i]=-\ka \,  K_i , &\qquad   [P_i,P_j]=\ka\, \frac 1{c^2}\, \epsilon_{ijk}J_k , &\qquad[P_0,J_i]=0  ,
\end{array}
\label{fa}
\ee
where $c$ is the speed of light; hereafter Latin indices run as $i,j,k=1,2,3$, Greek indices run as $\alpha=0,1,2,3$ and sum over repeated indices will be understood.

As we showed in~\cite{BGH2019}, a homogeneous space of the Lie group with Lie algebra (\ref{fa}), corresponding to the momentum space in the Snyder model, can be described in terms of embedding coordinates $(\s_4,\s_0,\s_1,\s_2,\s_3)$ in a 5D ambient manifold that satisfy the constraint 
\begin{equation}
\Sigma_\ka\,  :\ \s_4^2-\ka\, \s_0^2+  \frac{\ka}{c^2}\bigl( \s_1^2+
  \s_2^2+ \s_3^2 \bigr)=1 .
\label{fh}
\end{equation}
In terms of these coordinates, the   isometries generators on the curved manifolds read
\bea
&& P_{0}=\s_4
\,\frac{\partial}{\partial \s_0}+\ka\, \s_0\, \frac{\partial}{\partial \s_4} \, ,\qquad\ \, P_{i}=-\s_4
\,\frac{\partial}{\partial \s_i}+\frac{\ka}{c^2}\, \s_i\, \frac{\partial}{\partial \s_4} \,  ,  \nonumber\\[4pt]
&&
K_{i}=     \s_0
\,\frac{\partial}{\partial \s_i}+\frac{1}{c^2}\, \s_i\, \frac{\partial}{\partial \s_0} \,  ,\qquad 
J_i=     - \epsilon_{ijk}\, \s_j \, \frac{\partial}{\partial \s_k}  \, .
\label{fj}
\eea
The Snyder--Lorentzian noncommutative spacetime is then obtained  upon the identification
\be
x^0:=\frac 1 c\,P_0,\qquad x^i := - c\,P_i ,\label{LorentzSnyderCoordinates}
\ee 
yielding
\be
\left[x^0,x^i\right]=\ka \, K_{i} , \qquad \left[x^i,x^j\right]= \ka\,\epsilon_{ijk} J_k ,
\ee
so that spacetime coordinates inherit the noncommutativity from the curvature of the (Anti)-de Sitter manifold.\footnote{Again, note that the model should be defined quantum-mechanically \cite{BGH2019}, however this is not important for the scopes of these notes.}

The full Snyder--Lorentzian phase space is constructed by introducing momenta as objects that are commutative and transform as vectors under Lorentz transformations. As seen for the Snyder--Euclidean model, the choice is not univocal.  In the case  of Beltrami projective coordinates $(q_{0},q_{1},q_{2},q_{3})$, these come from the central stereographic projection in the 5D ambient space with pole  
$(0,0,0,0,0)$ of a point $ Q=(\s_4,\s_0,\s_1,\s_2,\s_3) \in \Sigma_{\ka}$  (\ref{fh}) onto   the projective hyperplane with $s_4=1$, namely
\begin{equation}
\s_4=\frac{1}{\sqrt{1-\ka\, q_0^2 +\frac{\ka}{c^2}(q_1^2+q_2^2+q_3^2)  }},\quad\
\s_\alpha=\frac{q_\alpha}{\sqrt{1-\ka\, q_0^2 +\frac{\ka}{c^2} (q_1^2+q_2^2+q_3^2)}},  \quad\  q_\alpha=\frac{\s_\alpha}{\s_4}  .
\label{fk}
\end{equation}
These   coordinates correspond to the physical momenta used in the original work by Snyder \cite{Snyder1947} via the rescaling \cite{BGH2019}:
  \be
 p_{0}:= c\,q_0,\qquad p_{i}:= \frac 1 c\, q_i  ,
\label{pa}
\ee
which are well-defined whenever 
\be
1+\ka \biggl( p_{1}^2+p_{2}^2+p_{3}^2  -\frac 1 {c^2}\, p_{0}^2\biggr)    >0 .
\label{gd}
\ee
Depending on the sign of $\ka$ this conveys the following momenta domain:
\begin{itemize}

\item  In the de Sitter Lorentzian case  with $\ka>0$,    the condition (\ref{gd}) yields 
\be
\frac 1 {c^2}\, p_{0}^2-(p_{1}^2+p_{2}^2+p_{3}^2)<\frac 1 \ka.
\label{domA}
\ee
For on-shell particles this   implies an upper bound on the mass
\be
c^2 m^2\equiv \frac 1 {c^2}\, p_{0}^2-(p_{1}^2+p_{2}^2+p_{3}^2),
\label{mass}
\ee
 given by
\be
c^2  m^2<\frac 1 \ka\,.
\label{domaindS}
\ee
\item
   In the  Anti-de Sitter Lorentzian case  with $\ka<0$ the condition (\ref{gd}) implies that
\be
p_{1}^2+p_{2}^2+p_{3}^2-\frac 1 {c^2}\, p_{0}^2<\frac 1 {|\ka|} .
\label{domB}
\ee
Hence for on-shell particles this leads to
\be
c^2  m^2>-\frac 1 {|\ka|}\,,
\label{domainAdS}
\ee
which is always satisfied.      
 \end{itemize}
 
 With the choice of spacetime coordinates (\ref{LorentzSnyderCoordinates}) and physical momenta (\ref{pa}) one obtains the following Snyder--Lorentzian phase space~\cite{BGH2019} (the choice $\ka<0$ was named the Anti-Snyder model in \cite{Mignemi:2011gr}):
 \be
\begin{array}{ll}
\left[x^0,x^i\right]= \ka \, K_{i} ,& \qquad \left[x^i,x^j\right]= \ka\,\epsilon_{ijk} J_k,  \\[2pt]
\displaystyle{ \left[x^0,p_{\alpha}\right]=\delta_{0\alpha}-\frac{\ka}{c^2} \, p_{0}p_{\alpha},}& \qquad {   \left[x^i,p_{0}\right]=\ka \, p_{0}p_{i},  }
\\[4pt]
 \left[x^i,p_{j}\right]=\delta_{ij}+\ka \, p_{i}p_{j},   &\qquad  \left[p_{\alpha},p_{\beta}\right]=0,  
\end{array}
\label{gb}
\ee
where the Lorentz boosts and rotations  take the usual form
\be
K_{i}= x^0 p_{i}+\frac 1 {c^2} \,  x^i p_{0} ,\qquad J_i= \epsilon_{ijk}  x^{j} p_k.\label{LorentzRepBeltrami}
\ee
The phase space variables in \eqref{gb} together with the generators $K_{i}$ and $J_{i}$ satisfy the Jacobi identities, indicating that the phase space is indeed invariant under Lorentz symmetries. Moreover, the spacetime coordinates and the momenta transform classically under boosts and rotations:
\be
\begin{array}{ll}
    [ J_i, x^0]=0, & \qquad  [ J_i, x^j]= \epsilon_{ijk}   x^k,\\[4pt]
      [ J_i, p_{0}]= 0, &\qquad  [ J_i, p_{j}]= \epsilon_{ijk}  p_{k} , 
  \end{array}
\label{vectorb}
\ee
\be
\begin{array}{ll}
\displaystyle{     [ K_{i}, x^0]=-  \frac 1{c^2} \,  x^i}, & \qquad \displaystyle{   [ K_{i}, x^j]=-  \delta_{ij}   x^0},  \\
\displaystyle{   [ K_{i}, p_{0}]    =    p_{i} },   &\qquad \displaystyle{   [ K_{i}, p_{j}]= \frac 1{c^2}\, \delta_{ij}  p_{0} }.
 \end{array}
\label{vectorc}
\ee


\subsection{Different choices of momenta in the Snyder--Lorentzian model}
 \label{31}

Similarly to what seen in the Euclidean case in section~\ref{21}, also the Snyder--Lorentzian model admits different choices of physical momenta, corresponding to different choices of coordinates on the (Anti-)de Sitter momentum manifold. 
Let us consider Poincar\'e projective coordinates  $(\tilde q_{0},\tilde q_{1},\tilde q_{2},\tilde q_{3})$ coming from  stereographic projection of a point $ Q=(\s_4,\s_0,\s_1,\s_2,\s_3) \in \Sigma_{\ka}$  (\ref{fh})  with   pole  $(-1,0,0,0,0)$ onto   the projective hyperplane with $s_4=0$. Then this projection is determined by  
\be
\s_4= \frac{1+\ka\, \tilde q_{0}^{2}- \frac{\ka}{ c^{2}}  (\tilde q_1^2+\tilde q_2^2+\tilde q_3^2)}{ {1-\ka\, \tilde q_{0}^{2}+\frac{\ka}{ c^{2}}(\tilde q_1^2+\tilde q_2^2+\tilde q_3^2)} }  ,\quad\  
\s_{\alpha}=\frac{2{ \tilde q_{\alpha}}}{ {1-\ka\, \tilde q_{0}^{2}+ \frac{\ka}{ c^{2}}  (\tilde q_1^2+\tilde q_2^2+\tilde q_3^2)}} ,\quad\     { \tilde q_{\alpha}}=\frac{\s_{\alpha}}{1+\s_4} .
 \label{PoincLorentz}
 \ee
 Physical momenta are then obtained from Poincar\'e   coordinates after the same rescaling  (\ref{pa}),
 \be
 \tilde p_{0}:= c\, \tilde q_{0},\qquad \tilde p_{i}:=\frac 1 c\, \tilde q_{i},
 \ee
 giving  rise to the following  Snyder--Lorentzian phase space  
   \be
\begin{array}{ll}
\left[x^0,x^i\right]= \ka \,   K_{i} ,& \qquad \left[x^i,x^j\right]= \ka\,\epsilon_{ijk}   J_k,  \\[2pt]
\displaystyle{ \left[x^0,\tilde p_{\alpha}\right]=\frac{\delta_{0\alpha}}{2}\biggl( 1+\frac{\ka}{c^{2}}\,\tilde p_{0}^{2}-\ka (\tilde p_1^2+\tilde p_2^2+\tilde p_3^2)\biggr)-\frac{\ka}{c^2} \, \tilde p_{0}\tilde p_{\alpha},}& \qquad  \left[x^i,\tilde p_{0}\right]=\ka \, \tilde p_{0}\tilde  p_{i}, 
\\[8pt]
\displaystyle{  \left[x^i,\tilde p_{j}\right]=   \frac{\delta_{ij}}{2}\biggl( 1+\frac{\ka}{c^{2}}\,\tilde p_{0}^{2}-\ka   (\tilde p_1^2+\tilde p_2^2+\tilde p_3^2 )\biggr)+\ka \, \tilde p_{i} \tilde p_{j}, }  &\qquad  \left[\tilde p_{\alpha},\tilde p_{\beta}\right]=0,  
\end{array}
\label{gb2}
\ee
where the Lorentz boosts and rotations    now adopt the form  
\be
K_{i}= 2\,\frac{x^0 \tilde p_{i}+\frac 1 {c^2} \,  x^i \tilde p_{0}}{1+\frac{\ka}{c^{2}}\tilde p_{0}^{2}-\ka  (\tilde p_1^2+\tilde p_2^2+\tilde p_3^2)} \, ,\qquad J_i= 2\,\frac{\epsilon_{ijk}  x^{j} \tilde p_k}{1+\frac{\ka}{c^{2}}\tilde p_{0}^{2}-\ka  (\tilde p_1^2+\tilde p_2^2+\tilde p_3^2)} \, ,
\label{pboost}
\ee
provided that
\be
\begin{array}{ll}
  \bigl[ x^0 \tilde p_{i}+\tfrac 1 {c^2} \,  x^i \tilde p_{0}, \tfrac{1}{c^{2}}\tilde p_{0}^{2}-   (\tilde p_1^2+\tilde p_2^2+\tilde p_3^2)\bigr] =0,\\[6pt]
 \bigl[ \epsilon_{ijk}  x^{j} \tilde p_k, \tfrac{1}{c^{2}}\tilde p_{0}^{2}-   (\tilde p_1^2+\tilde p_2^2+\tilde p_3^2)\bigr] =0 .
  \end{array}
\ee
It can be checked that the quadratic commutators (\ref{gb2}) satisfy  the Jacobi identities and that spacetime coordinates and   momenta are transformed as vectors under the Lorentz generators (\ref{pboost}) according to the expressions (\ref{vectorb}) and (\ref{vectorc}).

The Snyder--Lorentzian model resulting from the Poincar\'e projective coordinates  (\ref{gb2}) is related to the one resulting from the Beltrami coordinates (\ref{gb})  via the following change of coordinates in momentum space:
\be
\tilde p_{\alpha}=\frac{p_{\alpha}}{1+\sqrt{1-\frac{\ka}{c^{2}}\, p_{0}^{2}+ {\ka} (p_1^2+p_2^2+p_3^2)}},\qquad
 p_{\alpha}=\frac{2 \tilde p_{\alpha}}{ {1+\frac{\ka}{c^{2}}\, \tilde p_{0}^{2}- {\ka} (\tilde p_1^2+\tilde p_2^2+\tilde p_3^2)}}.
\ee

Clearly, other choices for projective coordinates, and thus for momenta, would  provide another expressions for the Snyder--Lorentzian phase spaces.


\subsection{Quasi-canonical structure of the Snyder--Lorentzian model}
 \label{32}
 
We have seen in section~\ref{22} for the Snyder--Euclidean model that one can make the phase space algebra diagonal by taking as physical momenta the ambient coordinates as given in \eqref{EuclideanAmbientMomenta}. Here we show that the same is true in the Lorentzian case. 
Indeed, by defining the physical momenta as
\be
\pi_{0}:= c \, s_{0},\qquad  \pi_{i}:= \frac 1 c \,s_{i},\label{QC}
\ee
  the commutators  (\ref{gb}) reduce to
 \be
\begin{array}{ll}
\left[x^0,x^i\right]= \ka \, K_{i} ,& \qquad \left[x^i,x^j\right]= \ka\,\epsilon_{ijk} J_k,  \\[4pt]
\left[x^\alpha,\pi_{\beta}\right]=\delta_{\alpha \beta} \sqrt{1+\frac{\ka}{c^{2}}\pi_{0}^{2}-\ka (\pi_{1}^{2}+\pi_{2}^{2}+\pi_{3}^{2})},  &\qquad  \left[\pi_{\alpha},\pi_{\beta}\right]=0 ,
\end{array}
\label{gbC}
\ee
with the Lorentz generators  being given by
\be
K_{i}=  \frac{x^0  \pi_{i}+\frac 1 {c^2} \,  x^i \pi_{0}}{\sqrt{1+\frac{\ka}{c^{2}}\pi_{0}^{2}-\ka (\pi_{1}^{2}+\pi_{2}^{2}+\pi_{3}^{2})}} ,\qquad J_i= \frac{\epsilon_{ijk}  x^{j} \pi_k}{\sqrt{1+\frac{\ka}{c^{2}}\pi_{0}^{2}-\ka (\pi_{1}^{2}+\pi_{2}^{2}+\pi_{3}^{2})}} .\label{gbL}
\ee
Notice that 
\be
\begin{array}{ll}
  \bigl[  x^0  \pi_{i}+\frac 1 {c^2} \,  x^i \pi_{0},    \frac{1}{c^{2}}\pi_{0}^{2}-  (\pi_{1}^{2}+\pi_{2}^{2}+\pi_{3}^{2}) \bigr] =0,\\[6pt]
 \bigl[ \epsilon_{ijk}  x^{j} \pi_k, \frac{1}{c^{2}}\pi_{0}^{2}-  (\pi_{1}^{2}+\pi_{2}^{2}+\pi_{3}^{2}) )\bigr]  =0.
  \end{array}
\ee

Of course, one can check that  the new momenta defined in  \eqref{QC} satisfy the required properties, since they are commutative and  transform as vectors under Lorentz transformations (as in  (\ref{vectorb}) and (\ref{vectorc})).
One can also use the map \eqref{QC} together with \eqref{fk}--\eqref{gd} to see that the following relation holds
\be
1+\ka (p_{1}^2+p_{2}^2+p_{3}^{2})-\frac{\ka}{c^{2}}\, p_{0}^{2} = \frac{1}{ 1+\frac{\ka}{c^{2}}\,\pi_{0}^{2}-\ka (\pi_{1}^{2}+\pi_{2}^{2}+\pi_{3}^{2})},
\ee
 thus leading to the   constraint (see (\ref{gd}))
\be
1+\ka (p_{1}^2+p_{2}^2+p_{3}^{2})-\frac{\ka}{c^{2}}\, p_{0}^{2}>0 \quad \Longleftrightarrow \quad 1+\frac{\ka}{c^{2}}\,\pi_{0}^{2}-\ka (\pi_{1}^{2}+\pi_{2}^{2}+\pi_{3}^{2})>0 .
\ee
Therefore, similarly to the Euclidean case, the domain of the new momentum space is interchanged between the de Sitter and   the Anti-de Sitter cases, that is
\be
\begin{array}{ll}
\mbox{Snyder--Lorentzian space with $\ka>0$:}\quad
&\pi_{1}^{2}+\pi_{2}^{2}+\pi_{3}^{2}-\frac{1}{c^{2}}\,\pi_{0}^{2}<\frac{1}{\ka}  .\nonumber\\[4pt]
\mbox{Snyder--Lorentzian space with $\ka<0$:}\quad  &\frac{1}{c^{2}}\pi_{0}^{2}-(\pi_{1}^{2}+\pi_{2}^{2}+\pi_{3}^{2}) <\frac{1}{|\ka|};
 \end{array}
\label{domain}
\ee
to be compared with (\ref{domA}) and (\ref{domB}).

We remark  that the  ambient coordinate $s_4$, not appearing in the definition (\ref{QC}), can be obtained from the constraint  $\Sigma_\ka$ (\ref{fh}) giving
 \be
  \s_4 \equiv \sqrt{ 1+\frac {\ka} {c^2}  \, \pi_0^2-\ka(  \pi_{1}^{2}+\pi_{2}^{2}+\pi_{3}^{2}) } \,  ,
\label{interpret}
 \ee
 which is just the term deforming the   commutator  $ [x^\alpha,\pi_{\alpha}]$ (\ref{gbC}).
 Moreover, this determines the time-like geodesic distance, say  $p_r$,   between the origin $O$ and a    point $Q$  on the 5D ambient space as~\cite{conf}   
 \be
 \s_4=\cosh(\sqrt{\ka}\, p_r) .
 \ee

Finally, we write the   power series expansion of the commutator $ [x^\alpha,\pi_\alpha]$   in terms of $\ka$:
\be
 [x^\alpha,\pi_\alpha]=   1+  \frac \ka {2}   \biggl( \frac 1{c^2}\, \pi_0^2- (    \pi_{1}^{2}+\pi_{2}^{2}+\pi_{3}^{2}) \biggr) + o[\ka^2] .
\ee
Consequently,  for a small value of $\ka$, the Snyder--Lorentzian phase space (\ref{gbC}) can be regarded as a quadratic algebra such that the $\ka$-deformation   is determined by the  square of the relativistic momentum 4-vector: $\frac 1{c^2}\, \pi_0^2- (    \pi_{1}^{2}+\pi_{2}^{2}+\pi_{3}^{2})$.


\section{Discussion}

The Snyder model provides an example of a noncommutative spacetime and deformed phase space that are compatible with standard Lorentz symmetries. The nontrivial commutator between spacetime coordinates is inherited from the commutator of translation generators over some underlying curved manifold, which is invariant under the required group of symmetries, namely either the de Sitter or the Anti-de Sitter manifold.
The full phase space is then constructed by requiring that momenta behave as vectors under  Lorentz transformations and are commutative.
These two conditions leave some freedom in the definition of momenta. 

In the original Snyder's work,  momenta were identified with the Beltrami projective coordinates of the curved (Anti-)de Sitter manifold. They were selected because the Lorentz symmetry generators have a standard representation in terms of  these momenta and the noncommutative spacetime coordinates, see eq. \eqref{LorentzRepBeltrami}. However, this simplification comes at the price of having a deformed phase space algebra with non-zero diagonal terms, eq. \eqref{gb}.

In this work we studied different possible choices of momenta, which turn out to be related to different choices of coordinates on the (Anti-)de Sitter manifold (we performed a similar analysis for the Snyder--Euclidean model, where space coordinates are identified with translation on a sphere/hyperboloid). In general, choices corresponding to some projective coordinates other than the Beltrami coordinates will result in a non-trivial representation of the Lorentz generators, see for example the representation using momenta corresponding to the Poincar\'e projective coordinates, eq. \eqref{pboost}. At the same time, the phase space algebra is not significantly simplified, see \eqref{gb2}.
A notable exception to this is found when the momenta are identified with the ambient coordinates of the (Anti-)de Sitter manifold, eq.  \eqref{QC}. In this case, while the representation of the Lorentz generator is still non-standard, eq. \eqref{gbL}, the phase space  \eqref{gbC} is diagonal. So these coordinates  provide a quasi-canonical structure for the Snyder model.  

We summarize the results just discussed in table~\ref{table1}, where we report both the Snyder--Euclidean and the Snyder--Lorentzian phase spaces and generators representations.

Of course our findings leave several questions open for investigations. Most importantly, we wonder about the phenomenological consequences of different choices of momenta. It might be possible that the effects of the modification of the phase space cancel out with the effects of the modification of the representation of symmetry generators. If this is not the case, what would be the most effective way to distinguish among the different models?


\begin{table}[htp]{\footnotesize
\caption{Snyder--Euclidean and Snyder--Lorentzian phase spaces with position and momenta operators $(x_\alpha, p_{\alpha})$ and $(x_\alpha, \pi_{\alpha})$.}
\label{table1}
 \begin{center}
\noindent
\begin{tabular}{l   l l l }
\hline
\\[-0.25cm]
 \multicolumn{4}{c}{Snyder--Euclidean phase spaces}\\[0.2cm]
\hline
\\[-0.25cm] 
\multicolumn{2}{l}{\quad $  \left[x_1,x_2\right]=  \kk \, J  $ } & \multicolumn{2}{l}{\quad $  \left[p_1,p_2\right]= 0  $  }\\[0.2cm]
\multicolumn{2}{l}{\quad $ \left[x_1,p_1\right]=1+\kk \, p_1^2 $ } & \multicolumn{2}{l}{\quad $   \left[x_1,p_2\right]= \kk\, p_1 p_2 $  }\\[0.2cm]
\multicolumn{2}{l}{\quad $\left[x_2,p_2\right]= 1+\kk \, p_2^2 $ } & \multicolumn{2}{l}{\quad $   \left[x_2,p_1\right]= \kk\,  p_1 p_2  $  }\\[0.2cm]
\multicolumn{2}{l}{\quad $ [x_{i},\pi_j]= \delta_{ij} \sqrt{ 1-\kk (\pi_1^2+\pi_2^2)}  $ } & \multicolumn{2}{l}{\quad $  \left[\pi_1,\pi_2\right]= 0  $  }\\[0.3cm]
\multicolumn{4}{c} { \quad $ \displaystyle{ J= x_1\, p_2-x_2\, p_1=  \frac  {x_{1}\, \pi_2-x_{2}\, \pi_1 }{ \sqrt{ 1-\kk (\pi_1^2+\pi_2^2)} }  }$ } \\[0.6cm]
\hline
\\[-0.25cm]
\multicolumn{4}{c}{Snyder--Lorentzian phase spaces}\\[0.2cm]
\hline
\\[-0.25cm] 
\multicolumn{2}{l}{\quad $\left[x^0,x^i\right]= \ka \, K_{i}  $ } & \multicolumn{2}{l}{\quad $   \left[x^i,x^j\right]= \ka\,\epsilon_{ijk} J_k  $  }\\[0.2cm]
\multicolumn{2}{l}{\quad $ \left[x^0,p_{\alpha}\right]=\delta_{0\alpha}-\frac{\ka}{c^2} \, p_{0}p_{\alpha} $ } & \multicolumn{2}{l}{\quad  $ \left[x^i,p_{0}\right]=\ka \, p_{0}p_{i} $ }\\[0.2cm]
\multicolumn{2}{l}{\quad  $  \left[x^i,p_{j}\right]=\delta_{ij}+\ka \, p_{i}p_{j} $} & \multicolumn{2}{l}{\quad $   \left[p_{\alpha},p_{\beta}\right]=0 $  }\\[0.2cm]
\multicolumn{2}{l}{\quad $ \left[x^\alpha,\pi_{\beta}\right]=\delta_{\alpha\beta} \sqrt{1+\frac{\ka}{c^{2}}\pi_{0}^{2}-\ka (\pi_{1}^{2}+\pi_{2}^{2}+\pi_{3}^{2})} $ } & \multicolumn{2}{l}{\quad $ \left[\pi_{\alpha},\pi_{\beta}\right]=0$  }\\[0.3cm]
\multicolumn{4}{c}{\quad $ \displaystyle{   K_{i}= x^0 p_{i}+\frac 1 {c^2} \,  x^i p_{0}=  \frac{x^0  \pi_{i}+\frac 1 {c^2} \,  x^i \pi_{0}}{\sqrt{1+\frac{\ka}{c^{2}}\pi_{0}^{2}-\ka (\pi_{1}^{2}+\pi_{2}^{2}+\pi_{3}^{2})}}   }$ } \\[0.5cm]
\multicolumn{4}{c}{\quad $ \displaystyle{  J_i= \epsilon_{ijk}  x^{j} p_k=\frac{ \epsilon_{ijk}  x^{j} \pi_k}{\sqrt{1+\frac{\ka}{c^{2}}\pi_{0}^{2}-\ka (\pi_{1}^{2}+\pi_{2}^{2}+\pi_{3}^{2})}}  }$ } \\[0.6cm]
 \hline
\end{tabular}
\end{center}
}
 \end{table}



\acknowledgments

This work has been partially supported by Ministerio de Ciencia, Innovaci\'on y Universidades (Spain) under grant MTM2016-79639-P (AEI/FEDER, UE), by Junta de Castilla y Le\'on (Spain) under grants BU229P18 and BU091G19. The authors acknowledge the contribution of the COST Action CA18108.

\end{document}